# Highlights

The outcomes were mainly highlighted on:

- Introduced for the first time a scheme to monolithically integrate selection switches for series or parallel switches and also their combinations to achieve a variety of inductance steps, resulting in digitally controlled variable inductance.
- Improved design for implementing low-profile and tall-profile 3D micro coils and transformers.
- Higher fill factor and more efficient usage of Si-wafer are attributed to the 3D micromachined structures.
- Introduced a design that increases the number of inductance steps by a factor equal to the number of micro-coils.
- Introduced a standard MEMS process and a 3D surface micromachining process to implement tall-profile coils with magnetic core materials, resulting in higher performance.
- Integration of Ni-based magnetic core to enhance the device performance.

# Design and Simulation of a Micro-coiled Digitally-Controlled Variable Inductor with a Monolithically Integrated MEMS Switch

A. Sharaf[1,2], Sh. M. Eladl[1], A. Nasr[1], M. Serry[2]

[1] Radiation Engineering Dept., NCRRT, Egyptian Atomic Energy, Cairo Egypt
[2] Dept of Mechanical Engineering, American University in Cairo, New Cairo, Egypt

**Abtract:**

This work introduces the design, analysis, simulation, and a standard MEMS fabrication process for a three-dimensional micro-coil with a magnetic core and a digital switch configuration using a completely integrated, fully MEMS-compatible process to achieve a digitally controlled inductance. The proposed design can also be utilized as a micro-transformer. The proposed design consists of five identical 3D coils and their corresponding MEMS switches. These coils are digitally controlled to achieve a variable inductor ranging from one-fifth of the coil inductance up to five times the coil inductance. A standard five-layer Polymumps process is proposed to fabricate the micro-coils and the integrated switches. Each micro coil is anchored directly on-chip, which is connected to the input signal from one side, and the other is connected to the switch. The Ni-based magnetic core improves the coil's response by confining and guiding the magnetic field in the magnetic device compared to Si core based by more than five times. The presented coil has the number of windings limited by the designed length and the minimum spacing that can be realized by standard optical lithography. The coil's diameter is also restricted by the limits defined by optical lithography, whereas the maximum height realizable by the Polymumps process limits the height of the magnetic core and accordingly results in lower inductor performance. Based on this technique, we present coils ranging from 100 $\mu$m length and ten winding up to 1000 $\mu$m length and 100 windings. The new monolithically integrated MEMS switches act as selectors to achieve a variable inductance with digital control to allow the selection among $n(n+1)/2$ inductance steps, where $n$ is the number of coils.

## 1. Introduction

Inductor, a passive electronic element, is one of the most important electronic elements along with other passive elements such as resistors, capacitors, and active elements such as diodes, transistors, and integrated circuits (IC) [1, 2]. It finds different applications in diverse areas ranging from filtering, tuning circuits, and oscillators up to complex RF circuits [3- 7]. Passive components such as inductors and transformers are crucial elements in switched-mode power supplies [8], biomedical devices [9, 10], microfluidics [11, 12], and magnetic sensors [13, 14]. Micro-coiled MEMS inductors offer significant advantages in these applications: the high quality factor and inductance. However, power MEMS inductors must have sufficient current handling capability, low electromagnetic interference, and low parasitic capacitance [15].

However, it remains the most complex element to be implemented using standard integrated circuit (IC) technology [16]. Although inductors are important as passive electronic elements in different application areas, such as sensors and actuators, their miniaturization has not been as successful as for resistors and capacitor passive electronic elements [17-19]. Thus, the MEMS community has directed a considerable amount of research effort to coil miniaturization during the past decades [20-22].

Microelectromechanical systems (MEMS) introduce many solutions for implementing diverse bulky physical sensors [23]. Due to its powerful metrics such as low cost, small size, mass production, low power dissipations, high performance, and compatibility with slandered IC technology fabrication processes [24, 25]. Many fields find MEMS technology a good solution that is appropriate to achieve all required aspects to implement different types of sensors [26, 27] efficiently. Many researchers work toward achieving low area, high efficiency, and fabrication compatibility with standard MEMS and IC fabrication processes [28- 31].

Three categories of monolithic MEMS coils were reported. Namely, planar spiral-shaped micro coils [32, 33], 3D solenoidal micro coils with rectangular cross-sections [34, 35], and solenoids with circular cross-sections [36, 37]. Micro coil emulates macro scale coil fabrication processes by focusing on achieving good characteristics such as a high level of integration, freedom in achievable coil inductance with and without magnetic core material, high quality factor, electrical winding insulation, high mechanical strength and stability, and low manufacturing and assembly costs [38-41].

A transformer is a magnetic sensor used in many applications to transform the voltage and current between two networks. The process is performed through two types of coils named primary and secondary coils. Implementing micro coils through MEMS technology opens the flour to implement micro-transformers efficiently. Generally, miniaturized transformers that use MEMS techniques have many potential advantages, such as low power consumption, low cost, high efficiency, high quality factor (QF), and high-frequency operation [42- 45].

This work introduces the design, analysis, simulation, and a standard MEMS fabrication process for implementing a tall-profile magnetic core, 3D micro-coil with a magnetic core and a digital switch configuration using a completely integrated, fully MEMS-compatible process to achieve a digitally controlled inductance. The proposed design provides reliability, mass production, and a high yield for the fabricated micro coils. The simulations show a standalone design for implementing micro coils through a standard Polymumps® process [46].

## 2. Proposed Design

The proposed design implements the micro-coil in two successive layers and connects the two layers through vias opened in the magnetic core layer. This design allows the building of individual coils with variable inductance and micro transformers.

The proposed design introduced here implements five identical coils and five identical single-ended cantilevers to select the appropriate coil in series, in parallel, or in a combination selection between series/parallel.

Figure 1-a shows the 3D layout of the proposed design. There are two anchors for electrical connections and mechanically support the micro coil to the substrate. The lower layer was deposited and patterned over the insulator to achieve electrical isolation from the substrate. This layer is fixed in the substrate. The magnetic core material is deposited and patterned over the lower layer. Through holes are opened in the magnetic core layer at predefined points to connect the top layer half winding of the coil to the lower layer half winding. Figure 1-b shows the side view of the micro-coil. All anchors and the lower layer are fixed to the substrate resulting in a minimized stress gradient effect on the micro-coil. Figure 1-c shows the elevation view of the micro-coil. As shown, it look-like a single-wound coil.

To achieve a variable inductance, switches are added to the proposed design. It depends on a single-ended micro cantilever design. These cantilevers are controlled across voltage to select the required value of inductance by combinations of series and/or parallel connections. Regarding the design simplifications, the cantilevers are made to vibrate in-plane to substrate instead of out-of-plane vibrations.

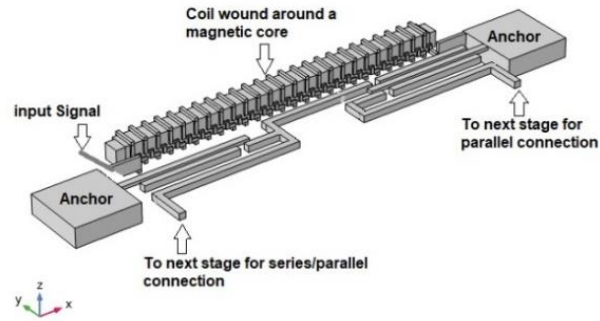

(a)

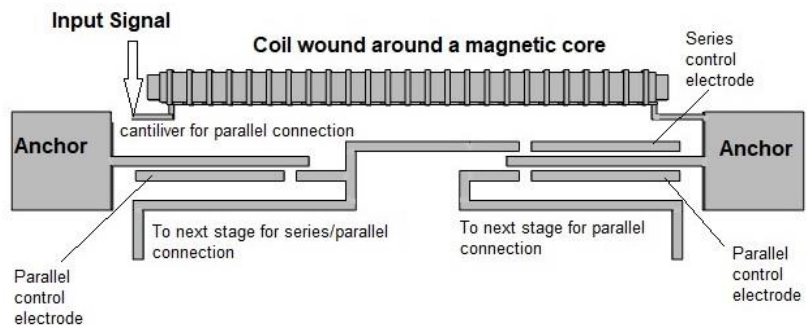

(b)

Figure 1: a) 3D layout of the proposed micro-coil along with the cantilever switch and connections (not to scale), b) The elevation view shows the micro-coil elements, anchors, cantilevers, control electrodes, and connections (not to scale).

Figure 1-b shows the top-view schematic diagram for the inductor along with the cantilever. The cantilever is a single-ended design fixed on one side, and the other is free to move. Two control electrodes are implemented on the two sides of the cantilever, one for series control and the other for parallel control. At the free tip of the cantilevers implemented, the electrical connection for the series and/or parallel connections. These configurations allow implementations of variable inductance ranging from one-fifth of a single coil inductance by connecting all cantilevers in a parallel configuration, up to five times the single coil inductance, in case of connecting the five coils in series.

The lumped element circuit diagram is presented in figure 2. The input signal is connected to the first coil ($L_1$) and, simultaneously, to the first parallel switch (the switch on the left of the diagram, namely, PSW1). The second terminal of the coil is connected to a parallel/ series switch (the switch on the right of the diagram, namely, PSSW1).

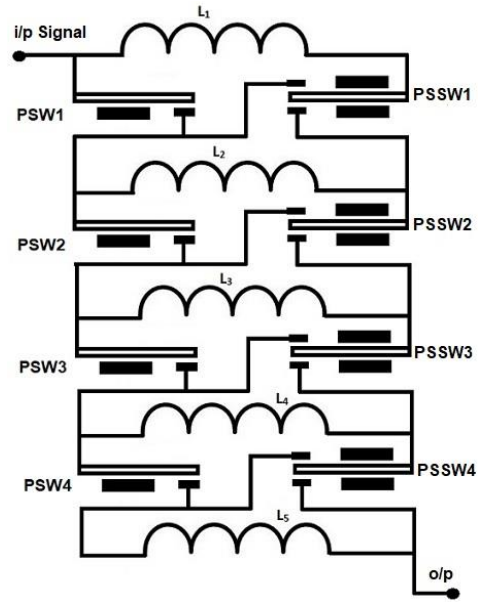

Figure 2: The lumped element model of the proposed circuit design for five coils case.

To connect a coil in a series configuration, then its parallel switch (PSW) should be disconnected at the same time the parallel/ series switch (PSSW) is closed so that its free tip is moved to the up position to select the first terminal of the next coil and so on. When connected in a parallel configuration, the parallel switch is closed so that the first coil terminal is connected to the first terminal of the next coil, and at the same time, the parallel/ series switch (PSSW) is closed so that it free tip moves to the down position to select the second terminal of the next coil and so on.

Introduced in Table 1 are the possible connection schemes for the five coils in both configurations and the resulting coil inductance. Figure 3 shows a bar graph for the total inductance of the proposed circuit as a function of the selection between series and parallel connections.

Table 1: different inductance values for different coils connection

| Step | No. of Series Switches | No. of Parallel Switches | Total Inductance |
|---|---|---|---|
| 1 | 0 | 5 | 0.2 L |
| 2 | 0 | 4 | 0.25 L |
| 3 | 0 | 3 | 0.33 L |
| 4 | 0 | 2 | 0.50 L |
| 5 | 1 | 0 | 1.00 L |
| 6 | 1 | 4 | 1.25 L |
| 7 | 1 | 3 | 1.33 L |
| 8 | 1 | 2 | 1.50 L |
| 9 | 2 | 0 | 2.00 L |
| 10 | 2 | 3 | 2.33 L |
| 11 | 2 | 2 | 2.50 L |
| 12 | 3 | 0 | 3.00 L |
| 13 | 3 | 2 | 3.50 L |
| 14 | 4 | 0 | 4.00 L |
| 15 | 5 | 0 | 5.00 L |

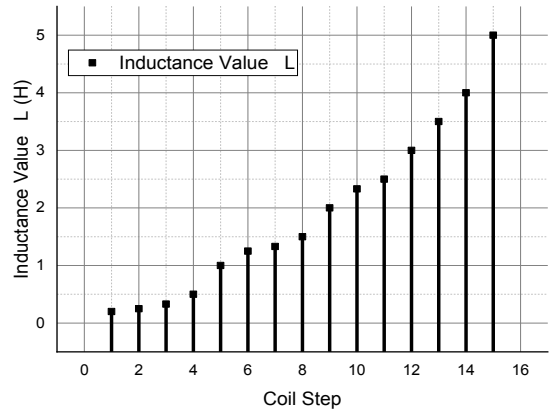

Figure 3: Represents the different values of total inductance according to the connected series and/or parallel switches for five coils case.

In general, the steps that can be obtained can be given by:

$$n + (n-1) + (n-2) + \cdots + (n-(n-1)) = n + (n-1) + (n-2) + \cdots + 1 = n(n+1)/2 \tag{1}$$

Where *n* number of coils. Thus the total number of steps can be estimated by:

$$Number\ of\ Steps = n(n+1)/2 \tag{2}$$

## 3. Results and Discussion

The proposed design was analytically analyzed. The proposed design is demonstrated and simulated using finite element analysis programs to verify the analytical results.

### 3.1. Theoretical Analysis

An inductor is basically just a coil of wire wound on a core which may be a magnetic material or air. According to Lenz's Law, the magnetic field will induce the motion of charges to resist change in the current; thus, the inductance is equal to the electromotive force (EMF) per unit rate of change of the induced current. Inductance can be calculated using the following formula, in the units of Henry (H):

$$L = \mu_o \mu_r \frac{N^2 A}{l}, \tag{3}$$

where *N* is the number of coil turns, *A* is the cross-sectional area of the coil, and *l* is the length of the coil, $\mu_o$ is a constant called the vacuum permeability constant and has the value $4\pi \times 10^{-7}$ H/m, $\mu_r$ is the relative permeability of the magnetic core used, and it is a material dependent value. Changing the core material of the coil will change the, results in change the coil inductance.

The designed parameters are summarized in Table 2. A variety of coil cross sections, lengths, and core materials used are supposed in the analysis to achieve good performance for the designed micro coil.

Table 2: Summarize the suggested design parameter for the proposed coil and its corresponding inductance.

| Area ($A$) (μm$^2$) | Length ($l$) (μm) | No. of turns ($N^2$) | Permeability ($\mu_o\mu_r$) | Core Cross Section area (μm$^2$) | Area on chip (μm$^2$) | Calculated inductance (nH) |
|---|---|---|---|---|---|---|
| 2×2 | 1000 | 225 | 30$\mu_o$ | 10×20 | 400×200 | 0.0339 |
| 2×3 | 1000 | 225 | 40 $\mu_o$ | 10×20 | 400×250 | 0.0678 |
| 2×4 | 1000 | 225 | 50 $\mu_o$ | 10×20 | 400×300 | 0.113 |
| 2×2 | 10000 | 15625 | 30 $\mu_o$ | 10×30 | 400×400 | 23.55 |
| 2×3 | 10000 | 15625 | 40 $\mu_o$ | 10×30 | 400×500 | 47.1 |
| 2×4 | 10000 | 15625 | 50 $\mu_o$ | 10×30 | 400×600 | 78.5 |

### 3.2. Finite Element Analysis

Finite element analysis using COMSOL Multiphysics version 5.5 was used to simulate the behavior of the coil and verify the analytical results. The cantilever was mechanically and electrically simulated to extract its mode shapes, frequency, and response to applied DC control voltages. The coil was simulated to extract the total magnetic energy produced with a 1A driving current.

#### 3.2.1. Mechanical Simulation

The mechanical simulation was performed on the moving parts of the design, namely the microcantilever. Both modal analysis, to extract the mode shapes and their resonances, and frequency response analysis, to determine the displacement values and directions of moving the free tip in response to applied DC control voltage and the required DC load voltage, are performed.

##### 3.2.1.1. Modal Analysis

Modal analysis is extracted to determine the mode shapes of the moving parts and their resonance frequency. Figures 4-a, 4-b, and 4-c, represent the fundamental mode shape at 32.772 kHz. Figures 4-d, 4-e and 4-f show the second mode shape at the resonance frequency of 81.848 kHz. It is clear that the second mode vibration is highly large in resonance than the fundamental mode and in shape is perpendicular to the substrate. The third mode shape with the resonance frequency of 205.33 kHz is shown in Figures 4-g, 4-h, and 4-i. The third mode vibrates parallel to the substrate like the fundamental but with a very high frequency over the fundamental mode. Diving the cantilever with the resonance frequency of the fundamental mode guarantees the proper operation of the design of both magnitudes of in-plane displacement and the proper connection of the desired coil configuration, namely parallel or series ones.

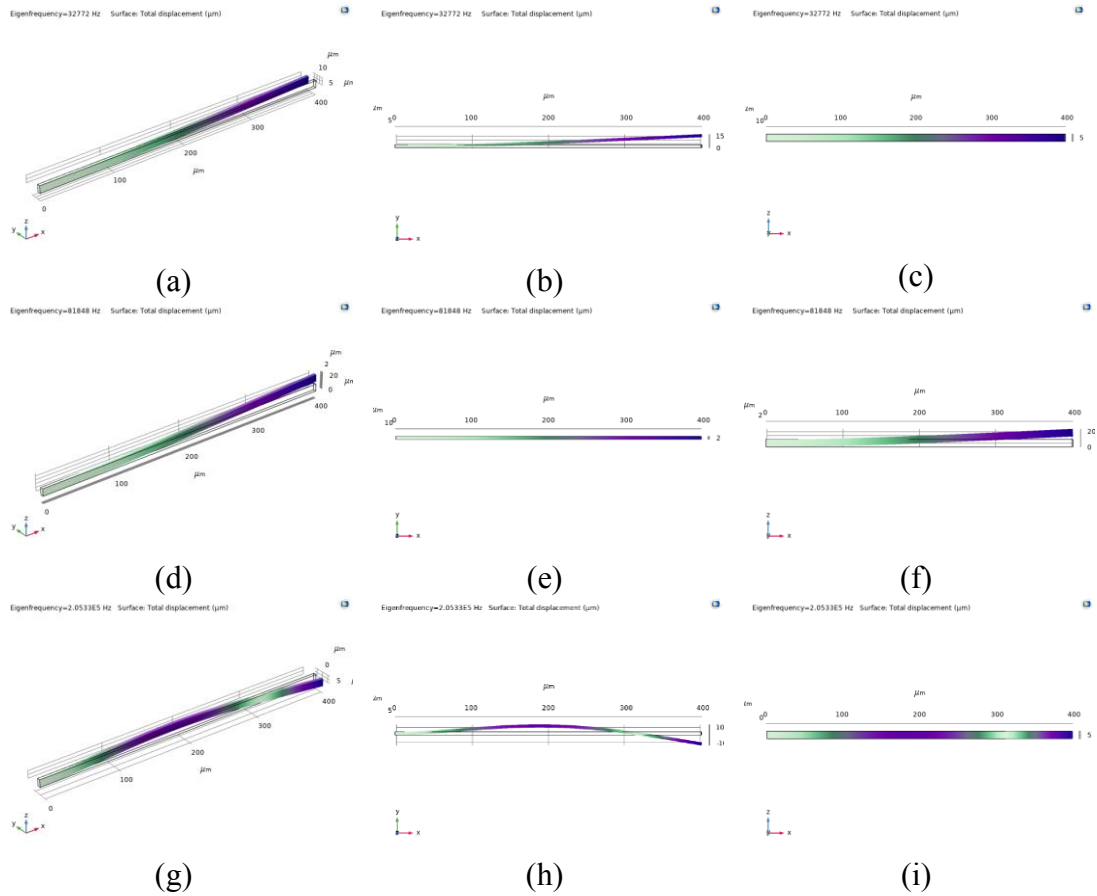

Figure 4: a) Fundamental mode shape at 32.772 kHz, moves right and left parallel to substrate; b) the top view of the first mode; c) the elevation view of the first mode shape; d) second mode shape at 81.848 kHz, moves up and down perpendicular to substrate; e) the top view of the second mode; f) the elevation view of the second mode shape; g) Third mode shape at 205.33 kHz, moves right and left parallel to substrate; h, the top view of the first mode; i) the elevation view of the first mode shape,

### 3.2.1.2. Frequency Response Analysis

Harmonic analysis was performed to investigate how much the design responds to the applied load and how it survives against load before it is destroyed. Figures 5-a and 5-b; show the moving cantilever's response to the static load effect. The diving voltage starting from 1 to 15 voltage was applied to the left control electrode. The cantilever responds to the applied voltage by left in-plane displacement. Figure 5-a represents the cantilever displacement at an applied voltage of 15V on the left control electrode, whereas figure 5-b represents the cantilever displacement for the right control electrode applied voltage of 15V. Figures 5-c and 5-d represent the corresponding elevation view for figures 5-a and 5-b, respectively.

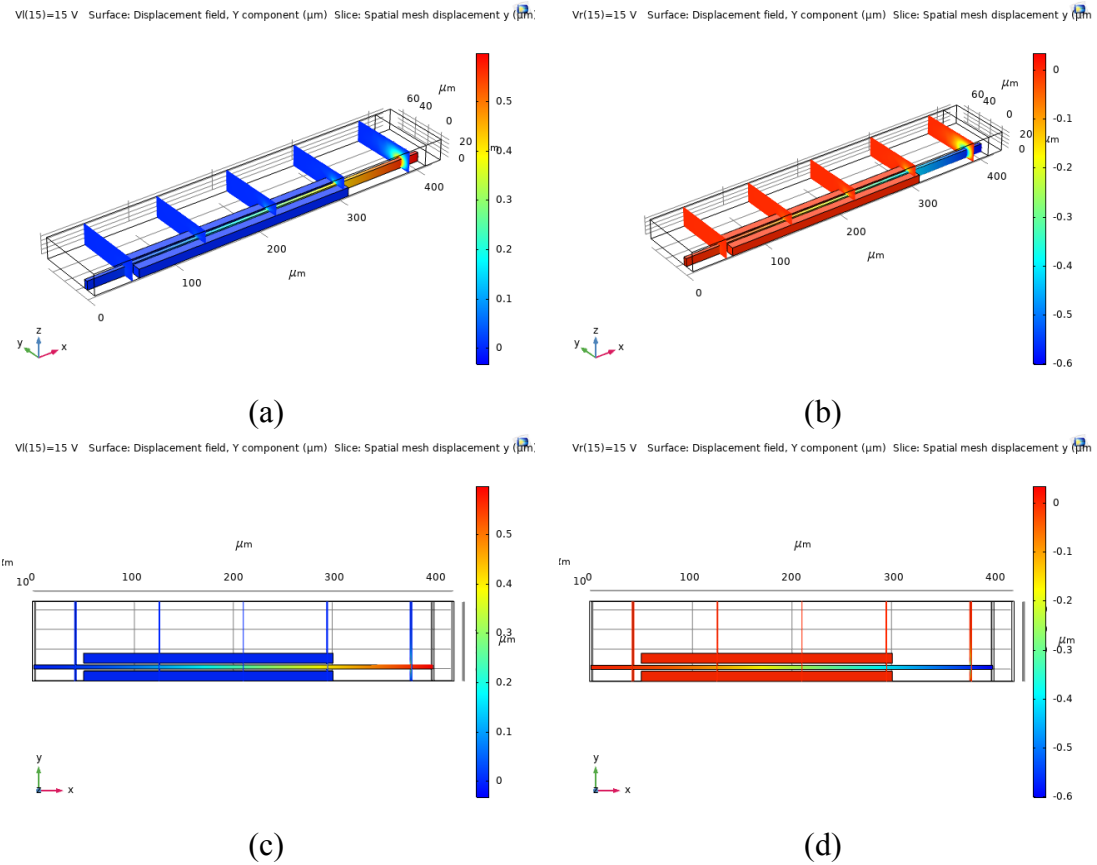

Figure 5: a) Cantilever displacement under 15V left actuation; b) cantilever displacement under 15V right actuation, c) XY-view for left actuation of the cantilever; d) XY-view for right actuation of the cantilever;

Figures 6-a; and 6-b; show the distribution of electric potential and electric field for the both cases introduces in figure 5. Figure 6-a shows the electric potential along the cantilever in response to the applied voltage on the left electrode form 1 to 15V. where figure 6-b represents the other case of right actuation.

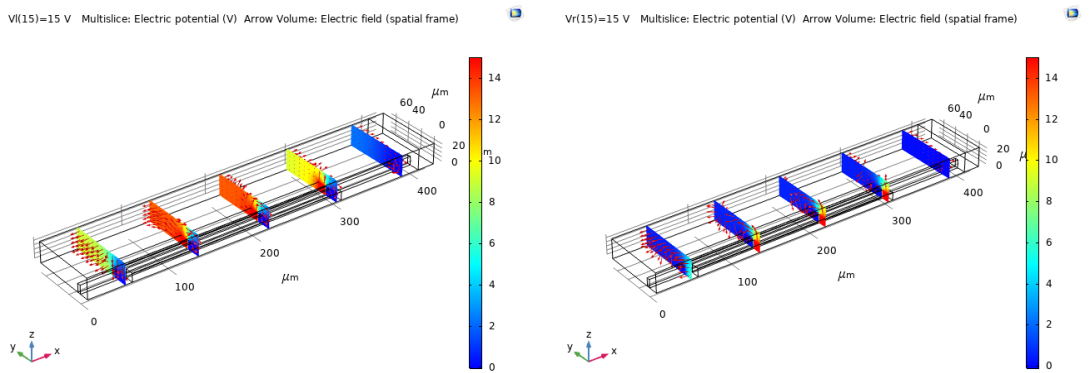

Figure 6: a) Cantilever electric potential and field distribution at 15V left actuation; b) Cantilever electric potential and field distribution at 15V right actuation.

Figures 7-a and 7-b; introduce the response of the cantilever beam to different applied voltages. It is clear that as the applied voltage increase, the beam displacement increase in both directions according to the applied control voltage. Figure 7-a represents the left actuation, while figure 7-b represents the right actuation.

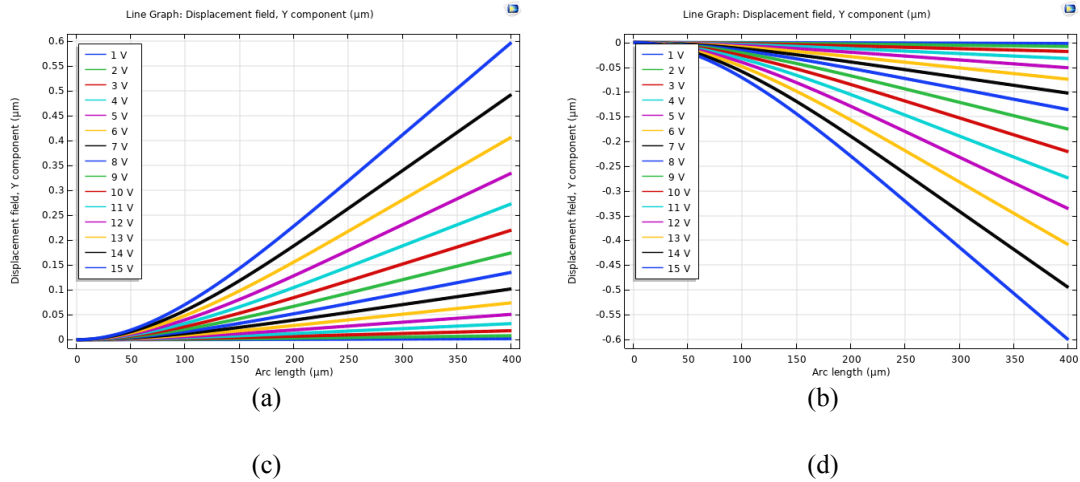

(a)                                    (b)

(c)                                    (d)

Figure 7: a) cantilever displacement under left actuation, and b) cantilever displacement under right actuation.

Figures 8-a; and 8-b; represent the beam displacement at pull-in voltage against frequency.

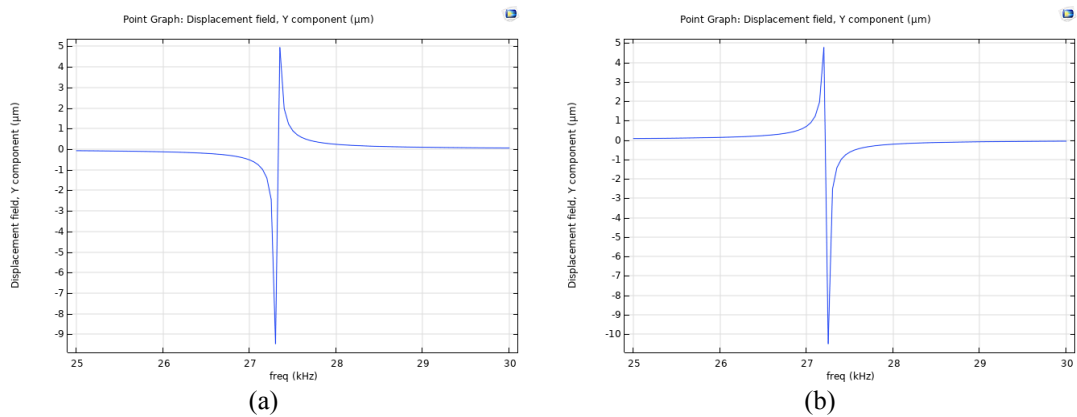

(a)                                    (b)

Figure 8: a) Represents the frequency response displacement of the free tip end of the cantilever under left actuation and b)Represents the frequency response displacement of the free tip end of the cantilever under right actuation.

Figures 9-a and 9-b; represent the static analysis for the point at the tip of the free end in response to the applied electric voltage on the control electrodes. Figure 9-a represents the left actuation, while figure 9-b represents the right actuation.

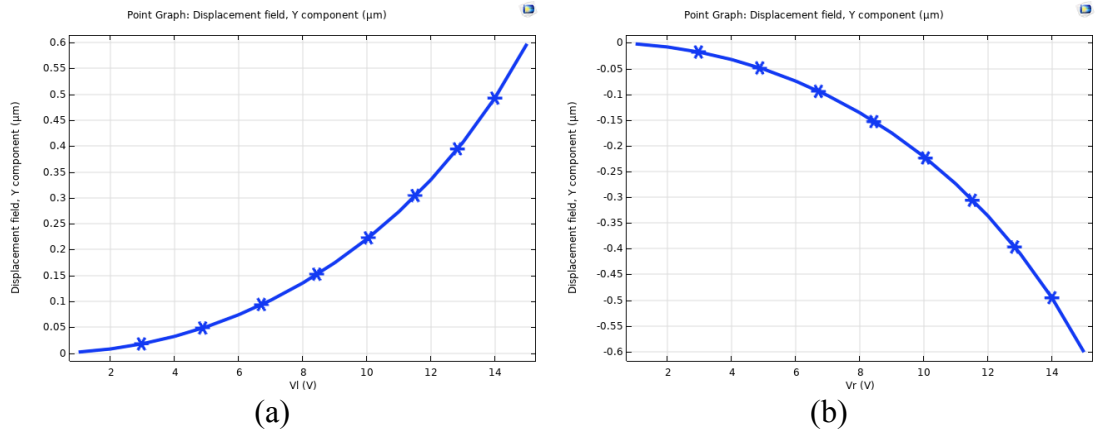

Figure 9: a) free tip end displacement under left actuation, b) free tip end displacement under right actuation

### 3.2.3 Inductor Design

Simulating the behavior of the designed coil to alternating current (AC) at high frequency is a milestone step in coil simulation. COMSOL multiphysics domain is required to link the various area concerning this topic. Structure mechanics, AC/DC, and RF domain physics are used to determine the behavior of the coil at a high-frequency range. The simulation performed takes on the part of the coil consisting of ten turns for simplicity. Figure 10 represents a magnetic core surrounded by ten turns of the proposed coil. Three different magnetic core dimensions and three different magnetic materials were simulated.

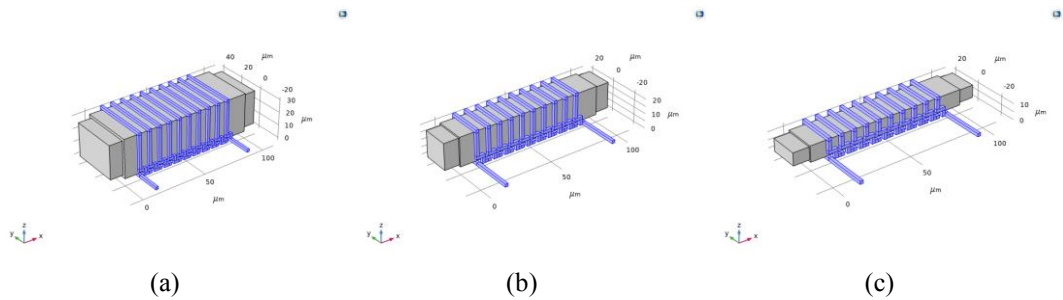

(a)　　　　　　　　　　(b)　　　　　　　　　　(c)

Figure 10: Represents a part of the coil with different dimensions' core used in the simulation.

Figure 11 shows the electric potential distribution along the 10 turns coil due to 1A excitation current. Due to the change in the core dimensions, the total length of the coil changed, resulting in different values of the electric potential distribution along the coil, as presented in Figures 11-a, 11-b, and 11-c.

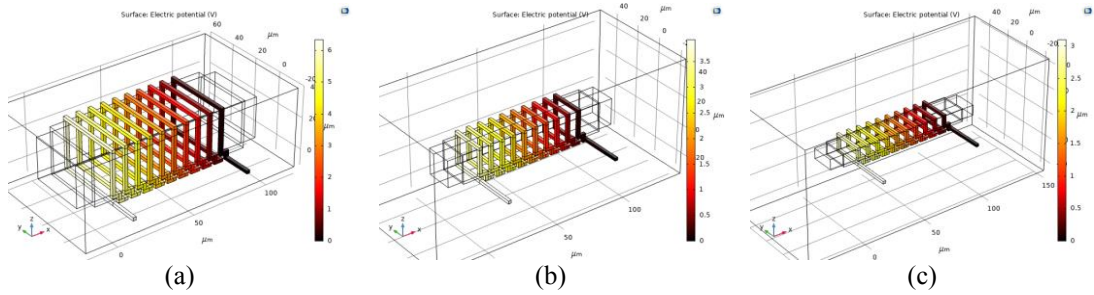

(a)            (b)            (c)

Figure 11: Represents electric potential of the coil with different dimensions' core used in this simulation.

Figure 12 simulates the total magnetic flux produced due to the 1A excitation current. The magnetic flux density for the same core dimensions differs significantly by changing the magnetic core material.

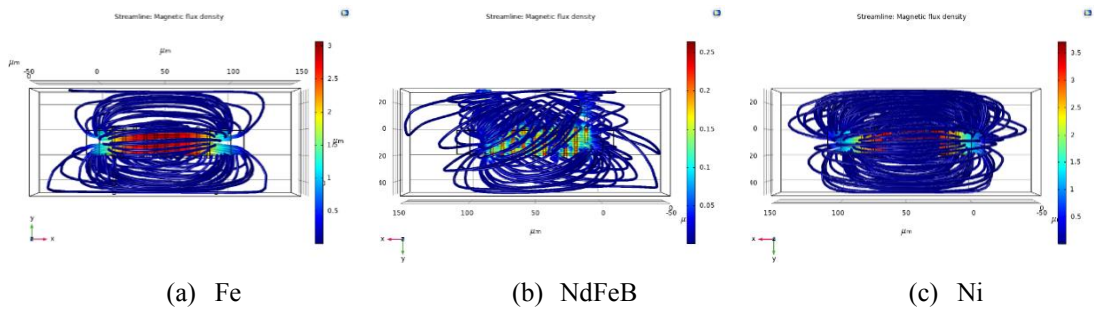

(a) Fe          (b) NdFeB          (c) Ni

Figure 12: Represents the magnetic flux density of the coil with 10 um thick core with different materials.

Figure 13 simulates the magnetic flux density for 20 um thick magnetic core with different magnetic materials. As the core thickness increased, the magnetic flux density increased for the same simulated magnetic material in the previous Figure 12.

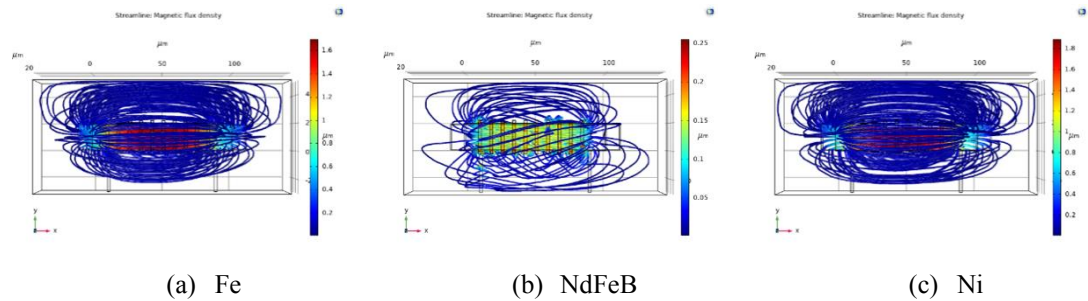

(a) Fe          (b) NdFeB          (c) Ni

Figure 13: Represents the magnetic flux density of the coil with 20 um thick core with different materials.

Figure 14 compares the magnetic flux density in the case of air core and with Fe and Ni core materials. Using magnetic material as a magnetic core increase the total magnetic flux density. Fe and Ni materials give magnetic flux densities relatively close to each other and greater than the air core case.

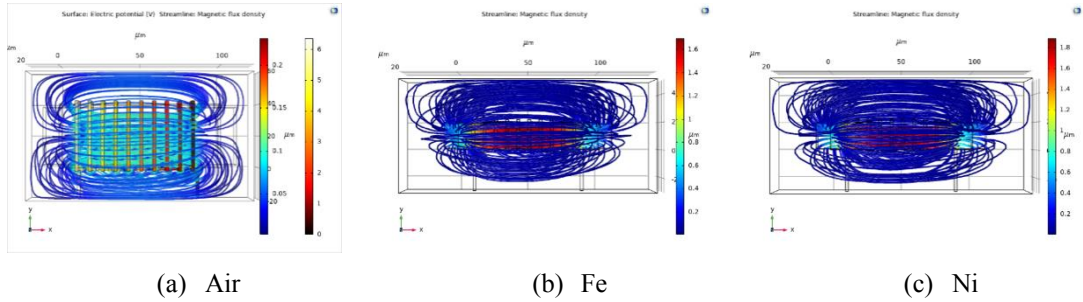

(a) Air        (b) Fe        (c) Ni

Figure 14: Represents the magnetic flux density of the coil with 20 um thick core with different materials.

Table 3 summarizes the evaluated inductance, resistance, and total energy extracted from the finite element analysis (COMSOL™). COMSOL™ evaluates the inductance using the current and the total magnetic energy. The inductance can be derived as the ratio of linked magnetic flux to current. For the case of a single coil, the inductance can also be calculated using current and magnetic energy. In particular, coil inductance can be expressed as:

$$L = \frac{2\,W_m}{I^2}, \tag{4}$$

where $W_m$ is the total magnetic energy in space. And *(I)* is the excitation current. Putting current equals 1A in the analysis, then twice the magnetic energy reflected the inductance value. This comparison is interpreted in figure 15.

Table 3: Inductance values for different core material and dimensions extracted from COMSOL simulations

| Core thickness (μm) | Core material | Inductance (nH) | Resistance (Ω) | Total Energy (nJ) |
|---|---|---|---|---|
| 10 | Fe | 3.845 | 3.103 | 1.923 |
| 10 | Ni | 4.624 | 3.101 | 2.312 |
| 10 | NdFeB | 0.5 | 3.101 | 0.2502 |
| 20 | Fe | 4.826 | 3.937 | 2.413 |
| 20 | Ni | 5.328 | 3.932 | 2.664 |
| 20 | NdFeB | 0.791 | 3.932 | 0.3957 |

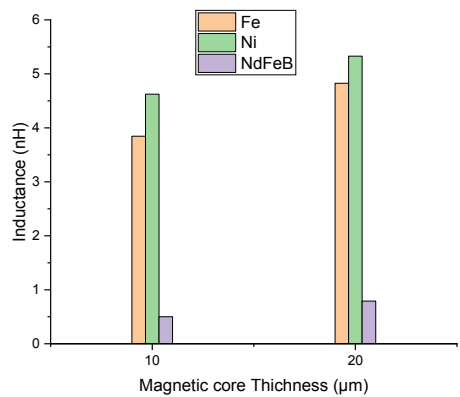

Figure 15: The effect of magnetic core material and thickness on the coil inductance.

## 4. Process Flow

Inductors consist of three materials: conductive material for windings, insulator material for electrical isolation between windings, and core material to concentrate the magnetic flux. The conductive material for windings is used to carry currents to produce the suitable magnetic flux. They are made of metals, such as copper (Cu). The insulator materials are dielectric thin films such as silicon oxide ($SiO_2$), aluminum oxide ($Al_2O_3$), or silicon nitride ($Si_3N_4$). The inductor cores can be made of air or magnetic materials.

Standard five-layer PolyMUMPs® can be used to implement the proposed inductor and its control switches. This process has the drawback of a lower aspect ratio, which results in lower magnetic core thickness. Accordingly limits the inductor performance. The process starts with an electrically isolated silicon wafer. The first step is to deposit and pattern another oxide layer. The patterned area in the next oxide to define the anchor area, control electrodes for series and parallel selections, the connection from one stage to the next inductor stage, and the groves for inductor lower connectors. The second step is to deposit the control electrodes, connection pads, anchors, and lower inductor layer. The third step is to deposit and pattern an oxide layer for electrical isolation for inductor winding and magnetic core. The fourth step is to deposit and pattern the magnetic core material. Then deposit the side wall and top inductor windings. Finally, the structure is released at cantilever parts. Figure 16 shows a simplified 2D layout for the proposed process.

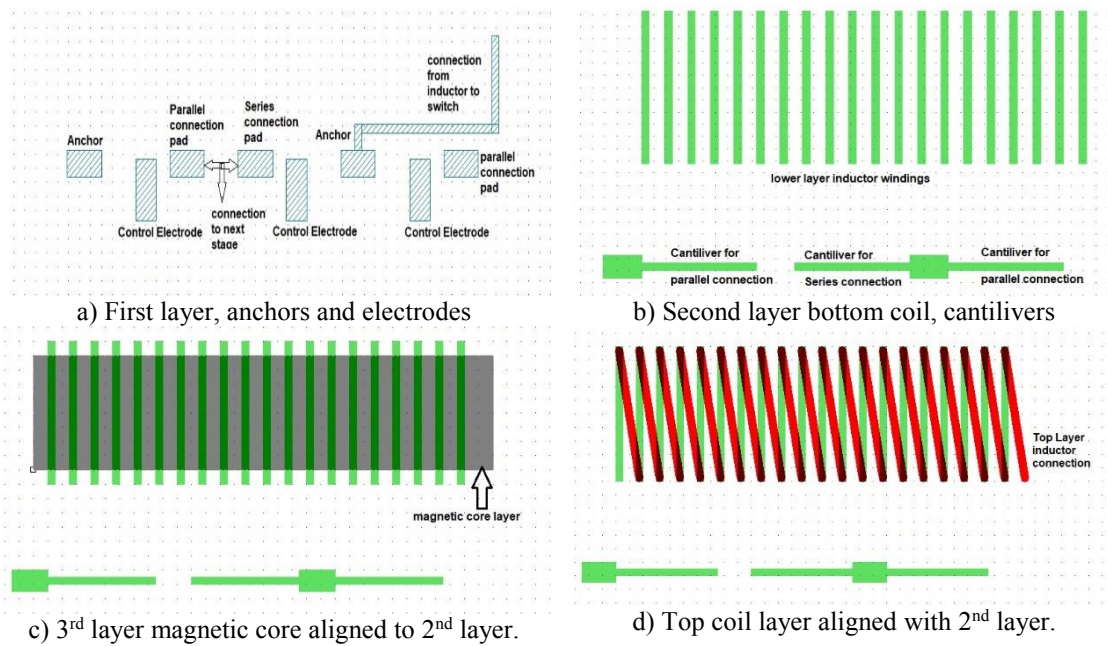

a) First layer, anchors and electrodes  
b) Second layer bottom coil, cantilevers  
c) 3rd layer magnetic core aligned to 2nd layer.  
d) Top coil layer aligned with 2nd layer.

Figure 16: summarized 2D process flow using PolyMUMPs.

3D surface micromachining technology is used to implement the 3D inductor introduced in this work. This process is suitable for building 3D micromachined inductors with low-profile air-core inductors, low-profile magnetic-core inductors, and tall-profile magnetic-core inductors [47, 48]

Figure 17 shows a summarized process flow. The process starts with an SOI wafer. This wafer has two benefits. First, the silicon layer is used to implement the cantilever switches, control electrodes, connection pads, and support anchors. The second benefit is using the insulator layer as a seed to build the 3D inductor. Then a layer of Cu is electroplating and patterned to form the bottom winding of the coil. Next, a thin insulator layer was deposited and patterned, followed by the deposition and pattern of magnetic core material. Then, a conductive material, Cu, is deposited and patterned to define the vertical and top windings of the inductor. Then, an insulator layer was deposited and patterned to offer the required isolation between windings. Finally, the structure was released at the cantilever part to define the required switches.

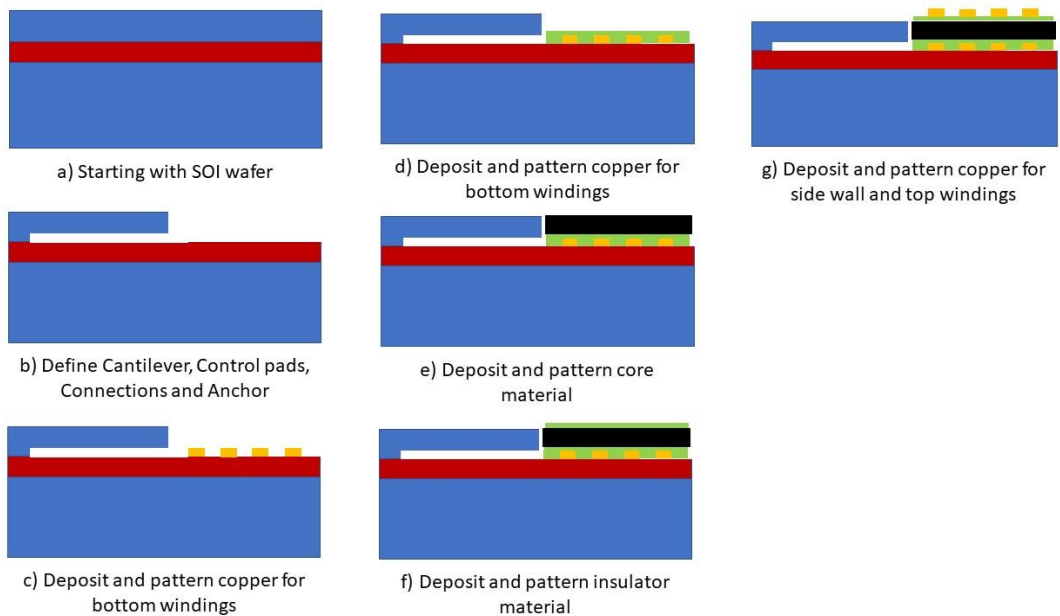

Figure 17: The proposed desin process flow using 3D surface micromaching foe tall profile magnetic core..

## 5. Conclusion

This work reported the design, analysis, and simulation of 3D micro-coils. The proposed micro coil introduced here was compatible with slandered MEMS fabrication processes. The main idea presented here is how to achieve a digitally controlled variable inductor. Five in-plane cantilevers were used to implement five switches to select among five coils. The selection can be parallel or series or a combination of parallel/series schemes to achieve a required inductance value. Depending on the fabrication technology, any required number of coil turns and dimensions can be implemented with a suitable magnetic core material. Fifteen inductance steps were obtained, ranging from 0.2L (H) up to 5L (H) can be achieved. The simulations ensure the designed idea. Both mechanical and electrical simulations confirmed and emphasized the success of the main idea behind the proposed design.